\documentclass[a4paper,fleqn,usenatbib]{mnras}
\usepackage{newtxtext,newtxmath}
\usepackage[T1]{fontenc}
\usepackage{ae,aecompl}
\usepackage{ulem,color}

\usepackage{graphicx}	
\usepackage{amsmath}	
\usepackage{subfigure}

\newcommand       \kpc		{\,{\rm kpc }}
\newcommand       \pc		{\,{\rm pc }}

\renewcommand{\sout}[1]{}
\hypersetup{draft}
\title{The Accelerating Pace of Star Formation}

\author[Caldwell \& Chang]{Spencer Caldwell$^{1}$\thanks{E-mail: caldwe44@uwm.edu} and Philip Chang$^{1}$
\\
$^{1}$Department of Physics, University of Wisconsin-Milwaukee, 3135 North Maryland Ave., Milwaukee, Wisconsin 53211, USA\\
}

\pubyear{2017}

\begin{document}
\label{firstpage}
\pagerange{\pageref{firstpage}--\pageref{lastpage}}
\maketitle
%
\begin{abstract}
We study the temporal and spatial distribution of star formation rates in four well-studied star-forming regions in local molecular clouds(MCs): Taurus, Perseus, $\rho$ Ophiuchi, and Orion A.  Using published mass and age estimates for young stellar objects in each system, we show that the rate of star formation over the last 10 Myrs has been accelerating and is (roughly) consistent with a $t^2$ power law. This is in line with previous studies of the star formation history of molecular clouds and with recent theoretical studies.
We further study the clustering of star formation in the Orion Nebula Cluster(ONC).  We examine the distribution of young stellar objects as a function of their age by computing an effective half-light radius for these young stars subdivided into age bins. We show that the distribution of young stellar objects is broadly consistent with the star formation being entirely localized within the central region.  We also find a slow radial expansion of the newly formed stars at a velocity of $v=0.17\,{\rm km\,s}^{-1}$, which is roughly the sound speed of the cold molecular gas. This strongly suggests the dense structures that form stars persist much longer than the local dynamical time.  We argue that this structure is quasi-static in nature and is likely the result of the density profile approaching an attractor solution as suggested by recent analytic and numerical analysis.
\end{abstract}

\begin{keywords}
  galaxies: star clusters: general -- galaxies: star clusters: individual -- galaxies: star formation -- stars: formation
\end{keywords}

\section{Introduction}

Per free-fall time, the mean star formation efficiency (SFE), the total mass of stars formed, in galaxies is of the order of a few percent and follows from the well characterized Kennicutt-Schmidt (KS) relations \citep{1989ApJ...344..685K,1998ApJ...498..541K} on galactic disk scales:
\begin{eqnarray}
  \dot\Sigma_* = \eta\Omega\Sigma_{\rm gas},
  \label{eqn: kennicutt_time}
\end{eqnarray}
where $\dot \Sigma_*$ is the surface density of star formation, $\Sigma_{\rm gas}$ is the total surface density of gas, $\Omega=v_c/R_d$ is the disk dynamical time, and $\eta\approx0.017$ is a dimensionless parameter that has an observationally determined value.

On smaller scales, ranging from $\sim1\kpc$ \citep{2010ApJ...722.1699S} in nearby galaxies, to $\sim100\pc$ in the Milky Way \citep{1988ApJ...334L..51M,1990ApJ...354..492M,1991ASPC...20...45E,2010ApJ...724..687L,2011ApJ...729..133M,2016ApJ...833..229L}, the simple KS relation breaks down. Here, these studies find relations similar in form to Equation (\ref{eqn: kennicutt_time}), but the dispersion of $\eta$ appears to vary with the scale on which the star formation is probed: measured values range from $\eta<10^{-3}$ to $\eta\approx0.5$.

What is responsible for the overall low efficiency of star formation on large scales and the large dispersion in $\eta$ on small scales?  Theoretical explanations for the low rate of star formation either invoke small scale physics, including magnetic fields \citep{1976ApJ...207..141M,1983ApJ...273..202S} or supersonic turbulence \citep{Padoan95,Krumholz+05}, or invoke large scale effects, including energy and momentum feedback from massive stars \citep{Murray+10}.  The variation of the star formation rate then arises from variations in the properties of molecular clouds, but it is presumed that equation (\ref{eqn: kennicutt_time}) continues to hold, i.e., the star formation efficiency is linear in time.

Recent theoretical \citep{2012MNRAS.420.1457H,2012ApJ...751...77Z,2014MNRAS.439.3420M,2015ApJ...806...31G,2015ApJ...800...49L,2015ApJ...804...44M,2017MNRAS.465.1316M,Murray+18} work of the physics of star formation in turbulent molecular clouds suggests that the KS relation (eq.[\ref{eqn: kennicutt_time}]) that is applicable on galactic scales is not merely scaled down for small scales.  Here rather than a linearly increasing SFE with time, the SFE appears to accelerate in time.  This acceleration has been explained either by global collapse of the cloud \citep{2012MNRAS.420.1457H,2012ApJ...751...77Z,2017MNRAS.467.1313V} or is the result of self-gravitational collapse of turbulent gas where the action of gravitational collapse feeds back on the turbulence providing some limited turbulent support against collapse \citep{2015ApJ...804...44M,2017MNRAS.465.1316M,Murray+18}.  In particular, \citet{2015ApJ...804...44M} showed that the density in collapsing turbulent gas approaches a fixed asymptotic profile while the velocity profile assumes a Keplerian profile, i.e., $v\propto M^{1/2} r^{-1/2}$ near the star giving rise to a mass accretion rate that scales like $t$ and hence a SFE that scales like $t^2$.   By suggesting an accelerating rate of star formation, these results echo an earlier observational result by \citet{1999ApJ...525..772P,2000ApJ...540..255P,2002ApJ...581.1194P}.

In this paper, we re-examine the results of \citet{1999ApJ...525..772P,2000ApJ...540..255P,2002ApJ...581.1194P} in light of recent theoretical results \citep{2015ApJ...800...49L,2015ApJ...804...44M}. We examine known stellar populations of Orion A, Taurus, $\rho$ Ophiuchi, and Perseus.  These four were chosen because of their well-studied stellar populations and availability of high quality observations.

We organize the paper as follows.  We briefly describe our methodology in \S~\ref{sec:Results} and immediately discuss our main results: a superlinear star formation rate (\S~\ref{sec:superlinear}) and a region of star formation that persists for many dynamical times (\S~\ref{ONC}).  In particular, we demonstrate that the observational results lend support to the turbulent collapse model of \citet{2015ApJ...804...44M}.  We discuss the results in the context of recent theoretical results in \S~\ref{discussion} and close with some conclusions in \S~\ref{sec:conclusions}.


\section{Methodology and Results}\label{sec:Results}

We focus on four star-forming complexes in this work: Orion A, $\rho$ Ophiuchi, Taurus, and Perseus.  Published values of the masses and ages of young stellar objects in these complexes were obtained and analyzed to study the temporal and spatial distribution of star formation.  The ages of stars were determined from the fitting of the observed luminosity and temperature to pre-main sequence models, but this fitting is subject to uncertainties.  For instance, deuterium abundance, accretion rate, and accretion geometry influence the "birth line" on the H-R diagram \citet{2006ApJ...641L.121T}.  Fortunately the birth-line mostly affects just $<1$ Myr ages, and does not alter the results dramatically.  The free-fall time used to determine the dynamical time of these complexes was obtained through the work done by \citet{2010ApJ...724..687L}.  This estimate of the density assumes that the clouds are spherical and may be an overestimate because not all clouds are spherical \citep{2012ApJ...745...69K}. 

\subsection{Superlinear Star Formation Efficiency}\label{sec:superlinear}

We first focus on the temporal distribution of star formation in Orion, Taurus-Auriga, Perseus, and $\rho$ Ophiuchi.
The Orion A molecular cloud is a substantial source of stellar activity due to the number of nebulae contributing to star formation; the most well-studied being the Orion Nebula Cluster.
Orion A is approximately 400 pc away \citet{2014ApJ...786...29S} and has a diameter of roughly 40 pc \citet{2016ApJ...818...59D}.  Orion A has a large stellar density of $2\times10^4/$pc$^3$ \citet{1999ApJ...525..772P}.  The total mass of the molecular gas in the cloud is 67,714 $M_{\odot}$ \citet{2010ApJ...724..687L}.

\begin{figure*}
	\centering
    \subfigure[Orion A]{%
		\includegraphics[width=\columnwidth]{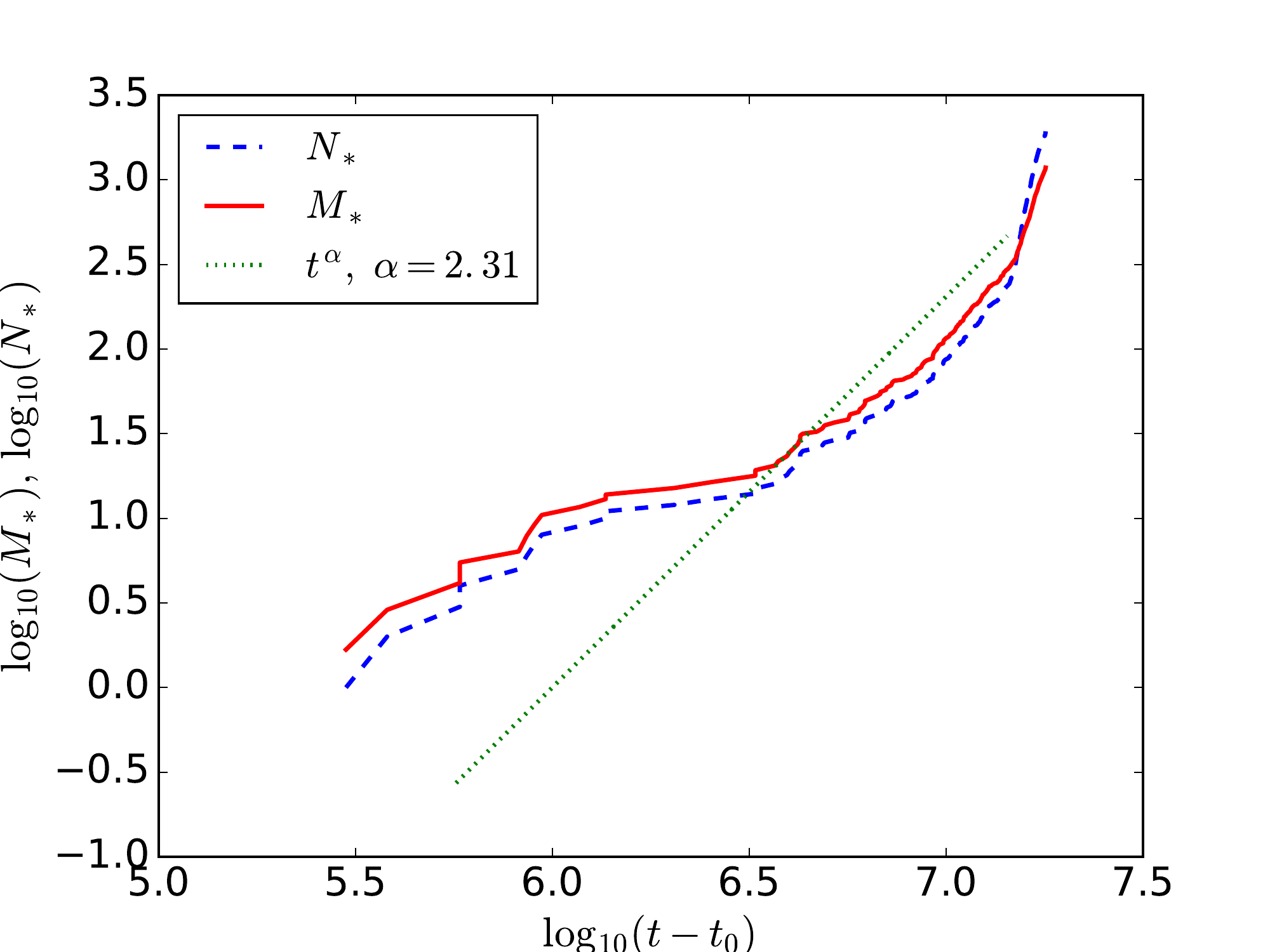}
		\label{fig:orion}}
	\subfigure[Taurus]{%
		\includegraphics[width=\columnwidth]{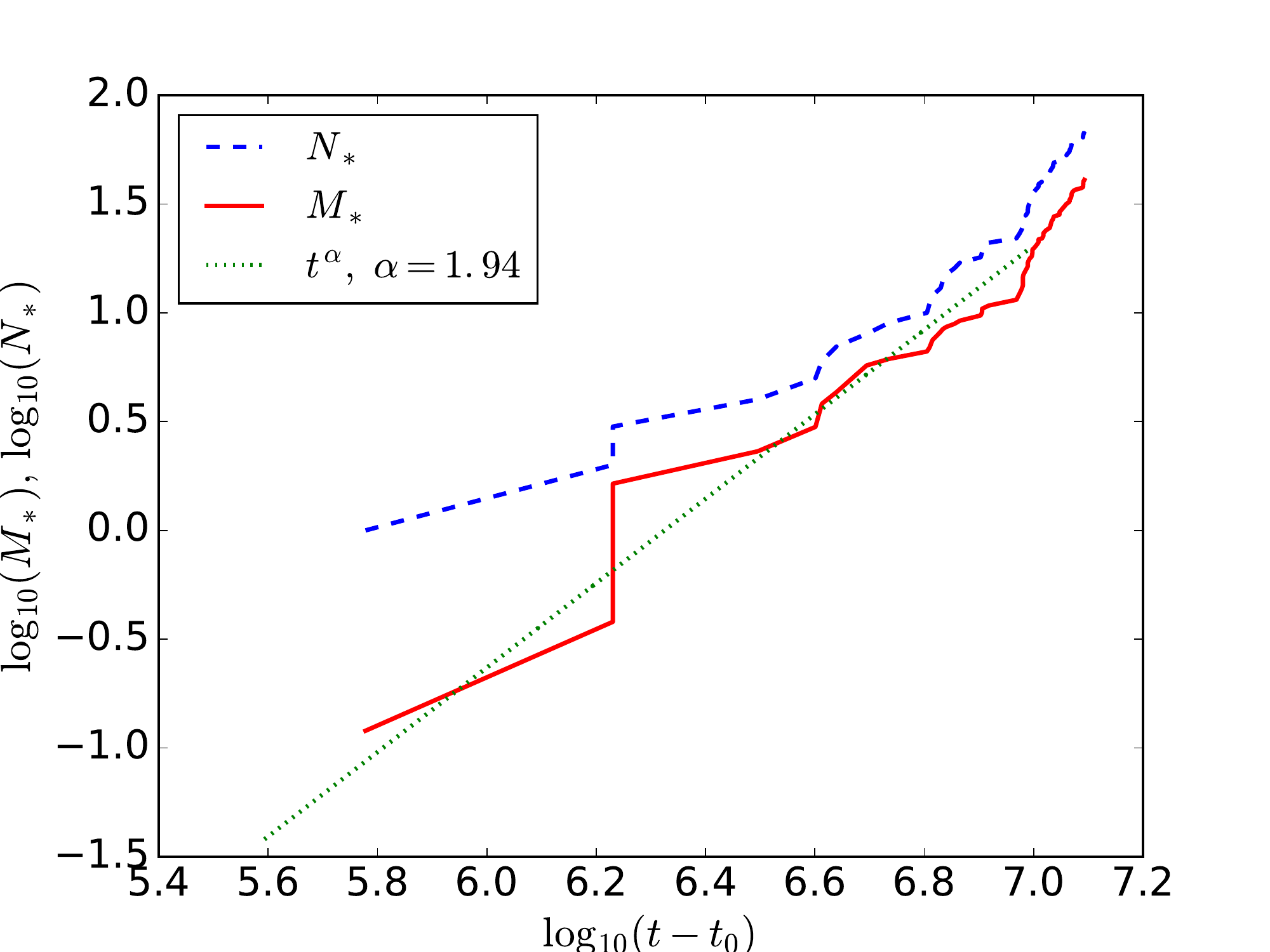}
		\label{fig:taurus}}
	\subfigure[Perseus]{%
		\includegraphics[width=\columnwidth]{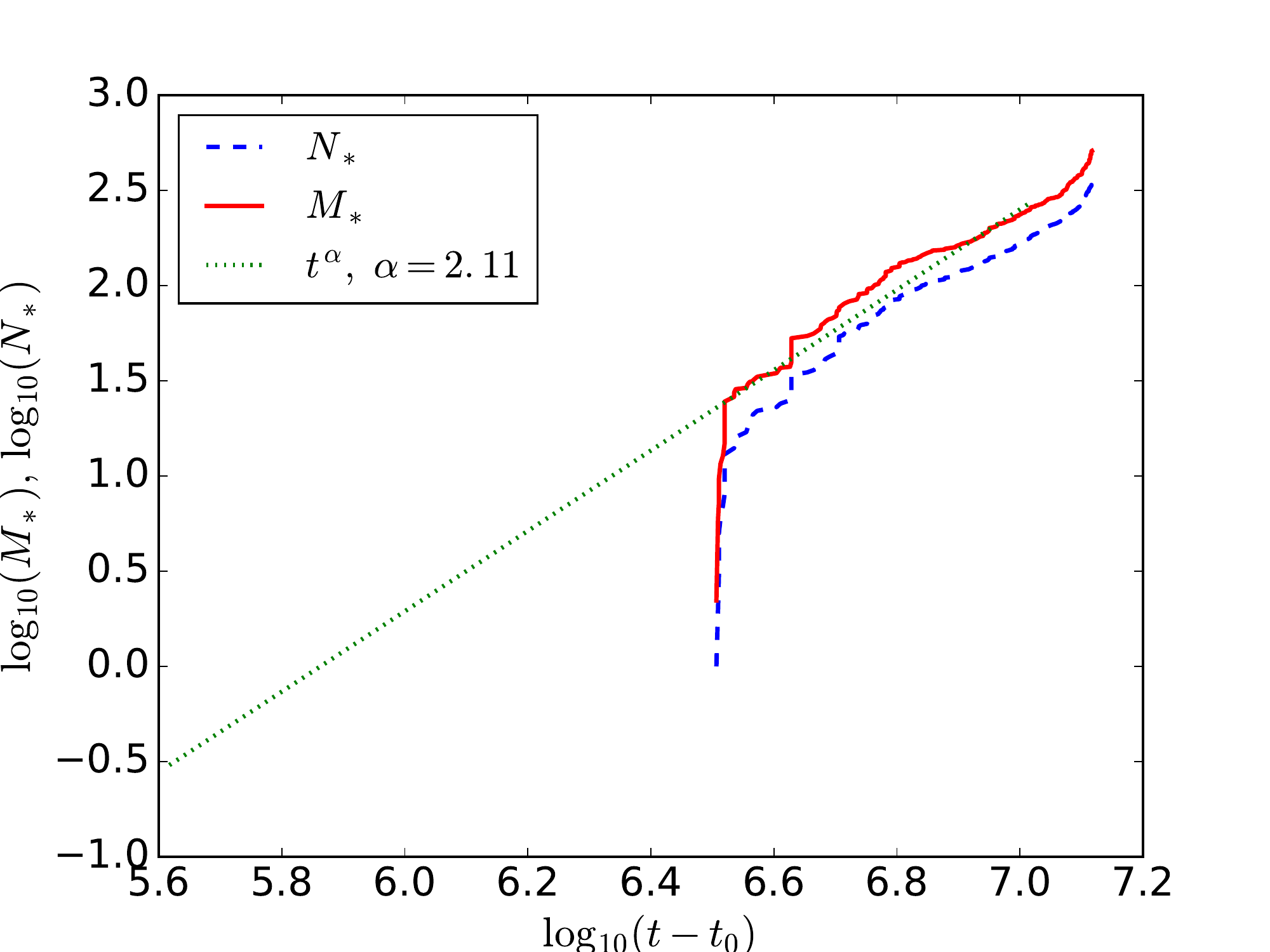}
		\label{fig:perseus}}
	\subfigure[$\rho$ Ophiuchi]{%
		\includegraphics[width=\columnwidth]{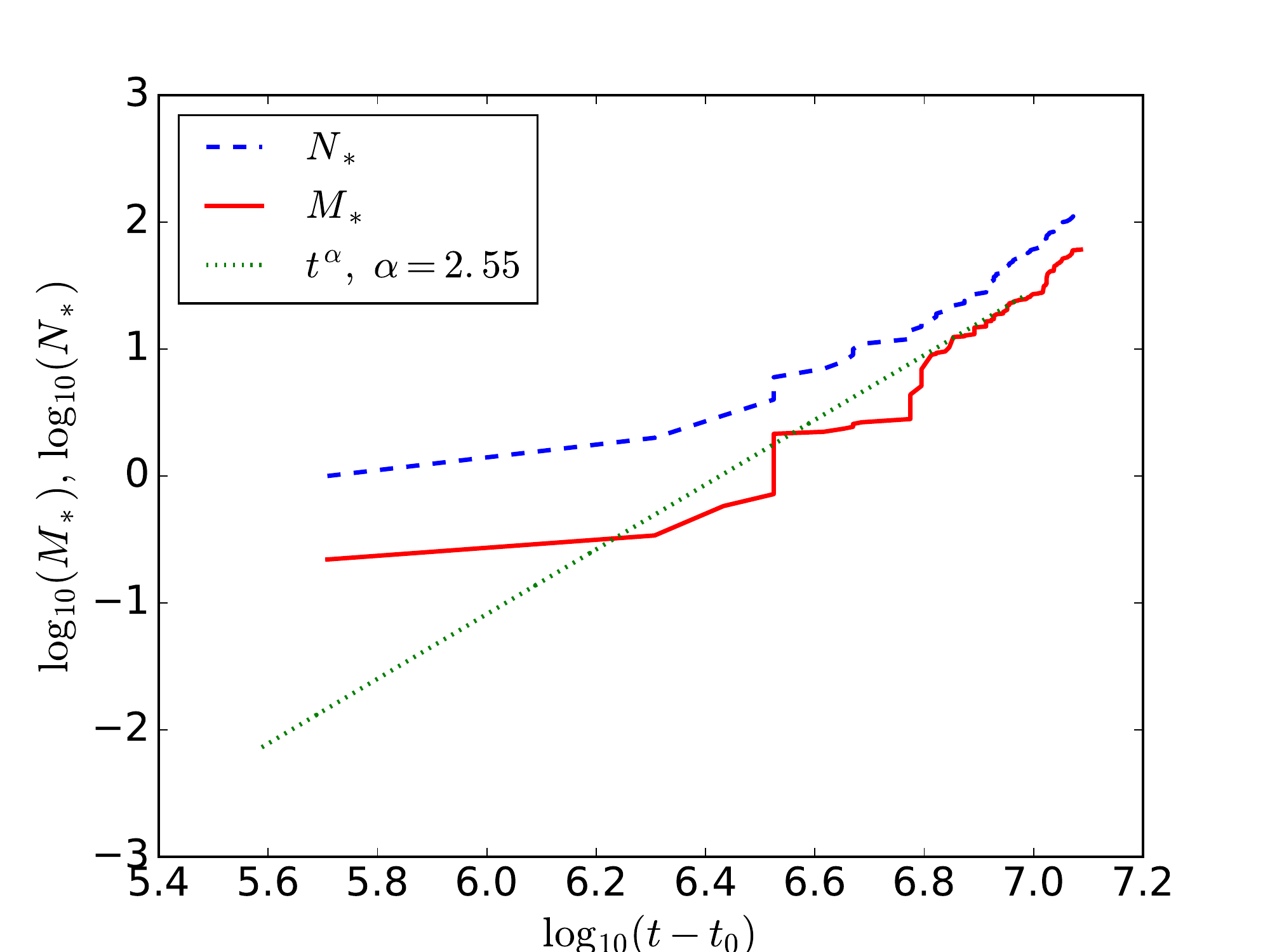}
		\label{fig:rho}}

	\caption{Total young stellar mass (red solid lines) and number of stars (blue dashed lines) formed since $t_0$ in Orion A (a), Taurus (b), Perseus (c), and $\rho$ Ophiuchi (d).  Here $t_0 = 2t_{\rm ff}$ in these plots.  Power law fits with time, $t$, are shown as a thin green dotted line and are fitted over the entire age range with the index given by $\alpha$, which ranges from $\sim 2 - 2.5$.}
	\label{fig:figure 1}
\end{figure*}

\citet{2016ApJ...818...59D} obtained data on 2691 stars in the Orion A Molecular Cloud.  We matched the data with confirmed members from \citet{2016yCat..18180059D}, yielding 2092 stars. Using the published mass and ages of these stars \citep{2016ApJ...818...59D} that were obtained using the evolutionary model of \citet{2000AA...358..593S}, we plot the SFE in the ONC as a function of time in Figure \ref{fig:orion} and find a power-law fit of $t^{2.31}$.


\citet{2010RMxAA..46..109K} provided mass and age estimations on 78 young, low-mass stellar objects (class 1 to class 3 sources) that were used to determine the star formation rate in the Taurus-Auriga complex.  Taurus is approximately 140 pc away \citet{1987ApJS...63..645U} and has roughly a 30 pc diameter \citet{2002ApJ...581.1194P}.  The cumulative mass of the molecular cloud constituents is 14,964 $M_{\odot}$ \citet{2010ApJ...724..687L}.  Far-infrared observations of these stars provided their spectral properties used for our analysis. The mass and age of the 78 young stellar objects members were calculated using the evolutionary model of \citet{Küçük1998}.  The mass and age estimates predicted by this evolutionary model were then plotted to show an increasing rate of star formation with a $t^{1.94}$ power law as seen in Figure \ref{fig:taurus}.

\citet{2015AJ....150...95A} gave age and mass estimates for 341 stars in Perseus, which is located around 300 pc from the sun \citet{2015AJ....150...95A} and has a cloud diameter of roughly 50 pc \citet{2008hsf1.book..308B}.  The total mass of the cloud is approximately 18,438 $M_{\odot}$ \citet{2010ApJ...724..687L}.  A WISE(Wide-Field Infrared Survey Explorer) survey was performed to identify young stellar objects in the Perseus complex and they also cross-matched the candidates with the SIMBAD database to obtain known young stellar objects \citet{2015AJ....150...95A}.  The ages and masses were given using the evolutionary model of \citet{2000AA...358..593S}. The star formation is accelerating with a $t^{2.11}$ power law shown in Figure \ref{fig:perseus}.

\citet{2011AJ....142..140E} offered mass and age estimates for 132 members of the stellar population of $\rho$ Ophiuchi, which is approximately a distance 130 pc from the sun \citet{2011AJ....142..140E} making it one of the closest star-forming MCs in our solar system.  The total mass of the cloud was found to be approximately 14,165 $M_{\odot}$ \citet{2010ApJ...724..687L}.  The multi-fiber spectrograph Hydra was utilized in obtaining spectra for the various members.  R- and I-band photometry were obtained with the 0.6m Curtis-Schmidt telescope to derive effective temperatures and bolometric luminosities for the stellar members.  The data obtained was compared with the evolutionary model of \citet{1997MmSAI..68..807D} to obtain the mass and age estimates seen in Figure \ref{fig:rho}.  The plot shows that the evolution in the $\rho$ Ophiuchi cloud is producing stars at a rate of $t^{2.55}$.

%
%

\begin{table}
	\centering
	\caption{Parameters and Power Law Fit indices of Orion A, Taurus, $\rho$ Ophiuchi, and Perseus.}
	\label{tab:SFR}
	\begin{tabular}{lccccr}
		\hline
		Name & $t_{\rm ff}$(Myr) &$ \alpha^a  $ &$ \alpha^b $ & $\alpha^c$  &References \\
		\hline
		Orion A &9.07 &1.79 & 2.31& 3.31 &  [1]\\
		Taurus& 6.22 &1.19& 1.94& 1.31 & [2] \\
		$\rho$ Ophiuchi &6.13& 1.09 & 2.55 & 1.52 &  [3] \\
		Perseus &6.55& 1.31& 2.11 &3.96 & [4]\\
		\hline
         \multicolumn{3}{l}{$^a$ $t_0 = t_{\rm ff}$, $^b$ $t_0 = 2t_{\rm ff}$,
        $^c$ $t_0 = 3t_{\rm ff}$}\\
        \multicolumn{6}{l}{[1] \citet{2016ApJ...818...59D} [2] \citet{2010RMxAA..46..109K}}\\
        \multicolumn{6}{l}{[3] \citet{2011AJ....142..140E} [4] \citet{2015AJ....150...95A}}
   	\end{tabular}

\end{table}

\begin{table}
	\centering
	\caption{Effective half-light radii of young stars in the ONC of various age bins}
	\label{tab:Dispersion}
	\begin{tabular}{ccc}
		\hline
		Age range(Myr)  &  $r_{\rm eff}$ (pc) & Number of Stars \\
        \hline
        $t_{\rm age}$< 0.5 &  0.66 & 159 \\
        0.5 <$t_{\rm age}$< 1.0  & 0.72 & 154\\
        1.0 <$t_{\rm age}$< 2.0  & 0.89 & 255\\
        2.0 <$t_{\rm age}$< 5.0  & 1.05 & 129\\
        \hline
	\end{tabular}
\end{table}

For the four molecular clouds discussed above, the SFE is superlinear as seen in Figure \ref{fig:figure 1}. This is the case for both number (dashed line) and total mass (solid line) of stars.  The fit of the SFE from the $t=t_0$ points is consistent with a power law of $\sim t^2$.  This is suggestive of the observed $t^2$ SFE in numerical simulations \citep{2014MNRAS.439.3420M,2015ApJ...800...49L,2015ApJ...806...31G,2017MNRAS.465.1316M,Murray+18} and found in analytic models of turbulent collapse \citep{2015ApJ...804...44M}.  This is in line with previous work by \citet{2000ApJ...540..255P}, who found that star formation in the ONC and other molecular clouds such as Taurus-Auriga, Lupus, Chamaeleon, $\rho$ Ophiuchi, Upper Scorpius, IC 348, and NGC 2264 have been accelerating with time, though they did not attempt to deduce the rate of acceleration.  Notably they argued that the collapse of dense independent clusters would not produce the observed SFE.  They instead argued that this acceleration can only be the result of a global process, which they attributed to global collapse.  Our quantification of this collapse and its association with a $t^2$ power law suggests that the turbulent (global) collapse of gas onto small star-forming regions is responsible for the observed acceleration.



\subsection{Localized Star Formation in ONC} \label{ONC}

Having examined the history of star formation in these MCs, we now examine the spatial distribution of star formation.  Of the four star-forming regions that we studied, only the ONC provides the combination of sufficient statistics and compactness to make definitive statements.  The other star-forming regions that we have studied have low statistics by comparison or have a star formation spread over a large area.  We will discuss these other regions in \S~\ref{discussion}.
\begin{figure}
	\centering
    \includegraphics[width=\columnwidth]{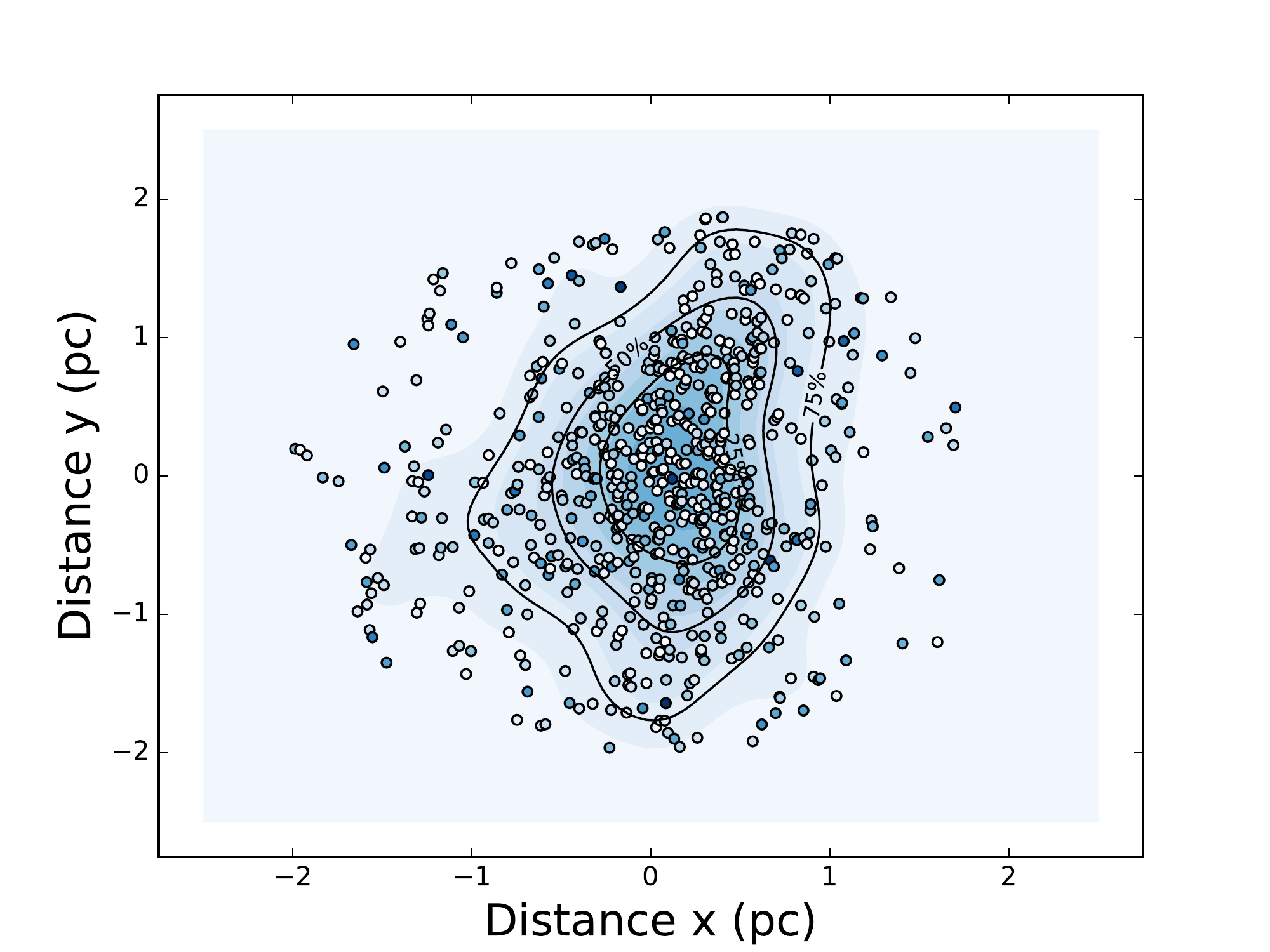}
    \caption{Positions of all the young stars within 2 pc (projected) of the center of the ONC, smooth density map, and contours that enclose 25\%. 50\%, and 75\% of the total light from the ONC.  Note the central concentration of the stars as evident from their position, peak of the smoothed intensity map, and contours.\label{fig:figure 2}}
\end{figure}

In Figure \ref{fig:figure 2} there is a clear increase in the density of stars toward the center. Figure \ref{fig:figure 2} shows 697 stars within 2 pc (projected) from the center of the ONC, which is determined from the centroid of the stars.  To quantify this, we smooth the position of the stars with a 2-D gaussian smoothing kernel to produce a smoothed density map, which is shown as an intensity map.  In addition, we plot contours that enclose 25\%, 50\%, and 75\% of the light of the stars.  In any case it is clear the distribution of stars peak toward the center.

\begin{figure*}
	\centering
    \subfigure[$t_{\rm age}$<0.5]{%
		\includegraphics[width=\columnwidth]{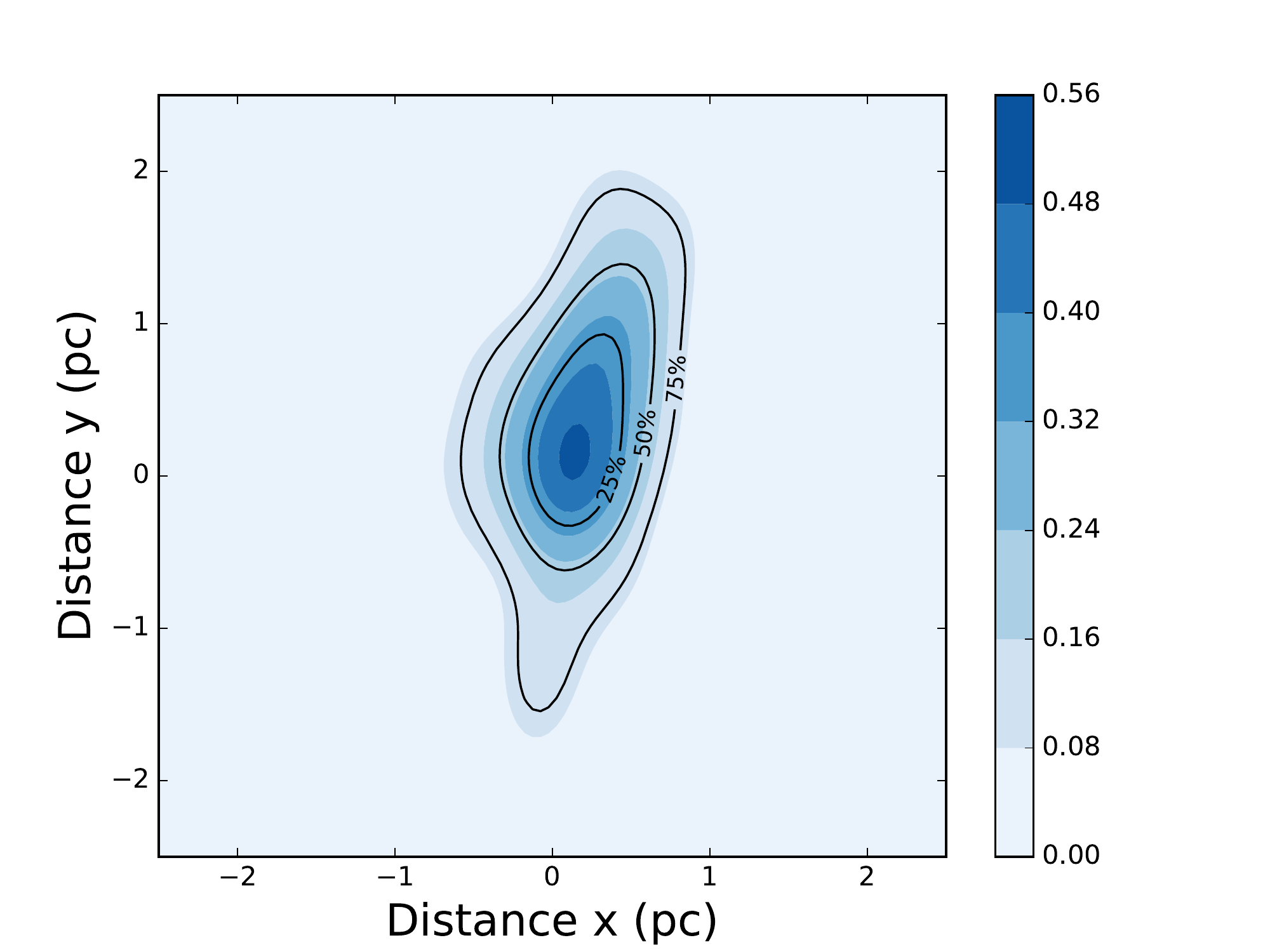}
		\label{fig:subfigure5}}
	\quad
	\subfigure[0.5<$t_{\rm age}$<1.0]{%
		\includegraphics[width=\columnwidth]{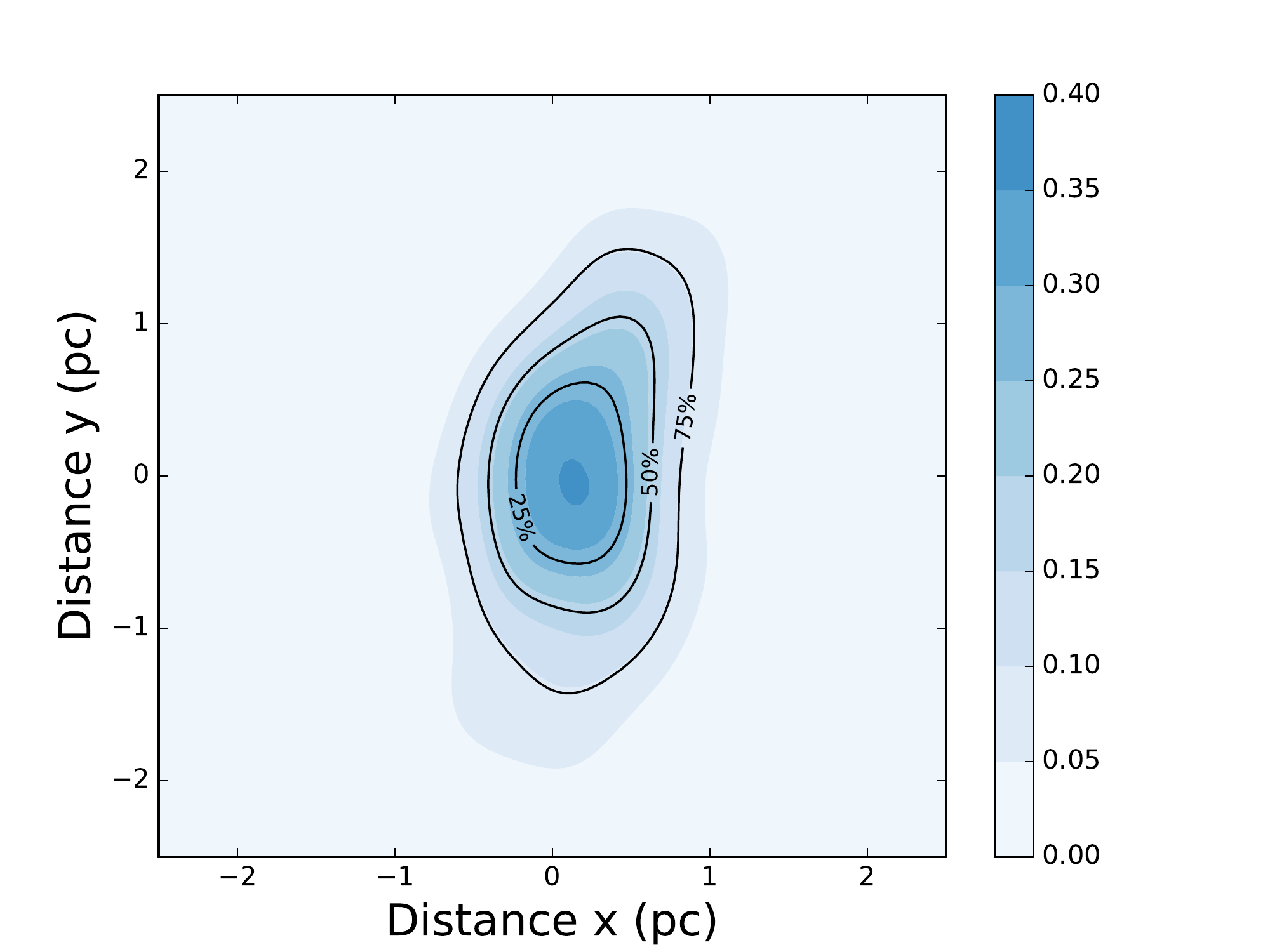}
		\label{fig:subfigure6}}
	\subfigure[1.0<$t_{\rm age}$<2.0]{%
		\includegraphics[width=\columnwidth]{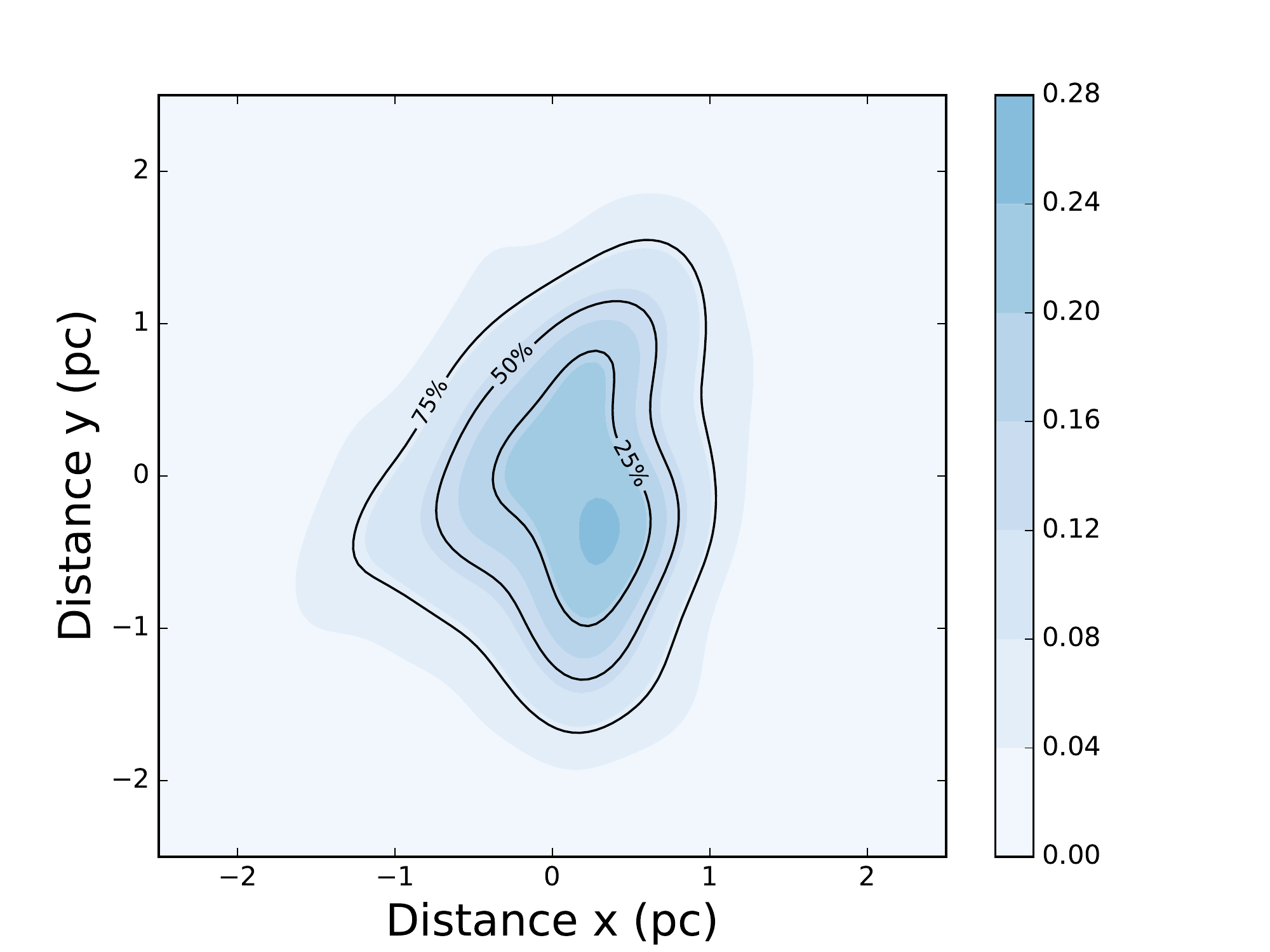}
		\label{fig:subfigure7}}
	\quad
	\subfigure[2.0<$t_{\rm age}$<5.0]{%
		\includegraphics[width=\columnwidth]{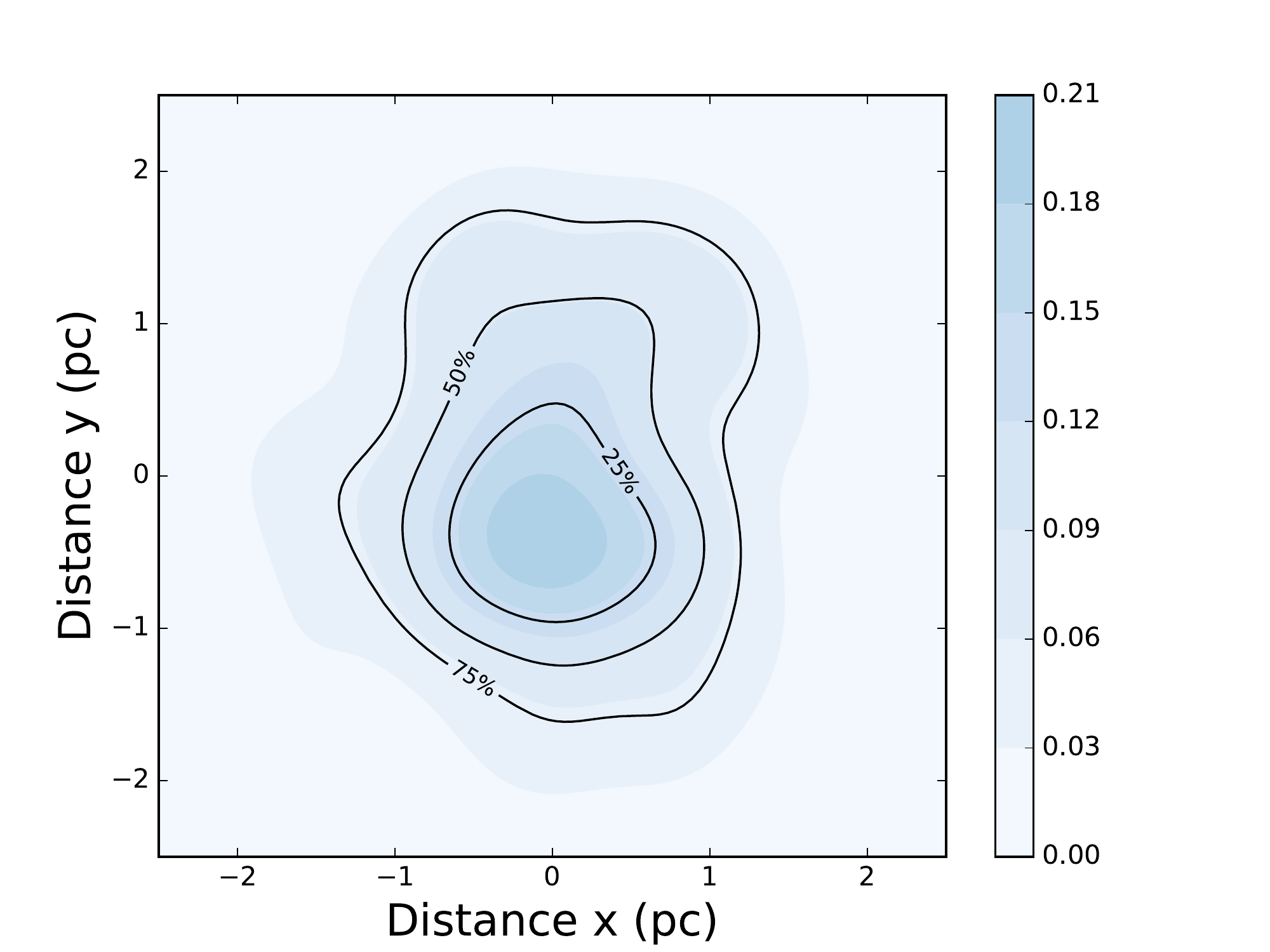}
		\label{fig:subfigure8}}

	\caption{Smoothed stellar density maps and contours for young stars in the ONC in different age bins.  The density maps are produced using a gaussian smoothing kernel of stellar distribution in different age bins.  The contours are traced at 25\%, 50\% and 75\% of the total light.  Note that the spatial distribution broadens as the stars age, a fact that is confirmed from calculating the effective half-light radius for the different bins.  $t_{\rm age}$ in Myrs.
	\label{fig:figure 3}}
\end{figure*}

Next, we break the stars into different age bins and compute the density maps and 25\%, 50\%, and 75\% contours for each bin in Figure \ref{fig:figure 3}.  Here the age bins are for $t_{\rm age}<0.5$ (plot a), $0.5 <t_{\rm age} < 1$ (plot b), $1 < t_{\rm age} < 2$ (plot c), and $2 < t_{\rm age} < 5$ (plot d), where the ages are in Myrs.  It is clear that younger stars are more concentrated toward the center.  To quantify this,  we note that the 50\% contour defines a region that contains half of the light.  We compute the area enclosed by this 50\% contour, $A_{50}$, and define an effective half-light radius,
$r_{\rm eff}$, by
\begin{eqnarray}
\pi r_{\rm eff}^2 = A_{50},
\end{eqnarray}
and list them in Table \ref{tab:Dispersion}.
The results of Figure \ref{fig:figure 3} and the effective radii computed in Table \ref{tab:Dispersion} show that younger stars are more centrally concentrated.  In addition, the different age bins all share a common center.  This is consistent with stars forming in the center of the ONC, which then migrate outward as they age. Moreover, this centralized star formation region has persisted for at least $5$ Myrs, which is over an order of magnitude longer than the central free-fall time of $0.35$ Myrs \citep{2006ApJ...641L.121T}.  This strongly suggests that, at least for the ONC, stars form in structures that persist over many local free-fall times.

In addition, stars appear to move outward as they age.  We can estimate the rate of this motion from the rate at which the effective radius moves outward.  This outward motion can be due to diffusion or the stars having some initial velocity.  The effect of diffusion is expected to be small as the relaxation time \citep{2008gady.book.....B} is much longer than the timescale of interest, i.e.,
\begin{eqnarray}
t_{\rm relax} \approx 0.1 \frac{N_{\rm stars}}{\ln \Gamma} t_{\rm cross} \approx 10 \left(\frac{N_{\rm stars}}{159}\right)\left(\frac{t_{\rm cross}}{0.7\,{\rm Myrs}}\right)\,{\rm Myrs},
\end{eqnarray}
where $N_{\rm stars}$ is the number of stars in the cluster, $t_{\rm cross}$ is the crossing time, and $\ln\Gamma$ is the Coulomb logarithm, which is taken to be order unity. Here, we take $N_{\rm stars}$ to be the number of stars in the central region, i.e., the centrally concentrated stars with $t_{\rm age} < 0.5$ Myrs.  We use the \citet{2006ApJ...641L.121T} value for the crossing time of $t_{\rm dyn}$ of $7 \times 10^5$ yrs for the ONC, i.e., twice the free fall time.

\begin{figure}
	\centering
    \includegraphics[width=\columnwidth]{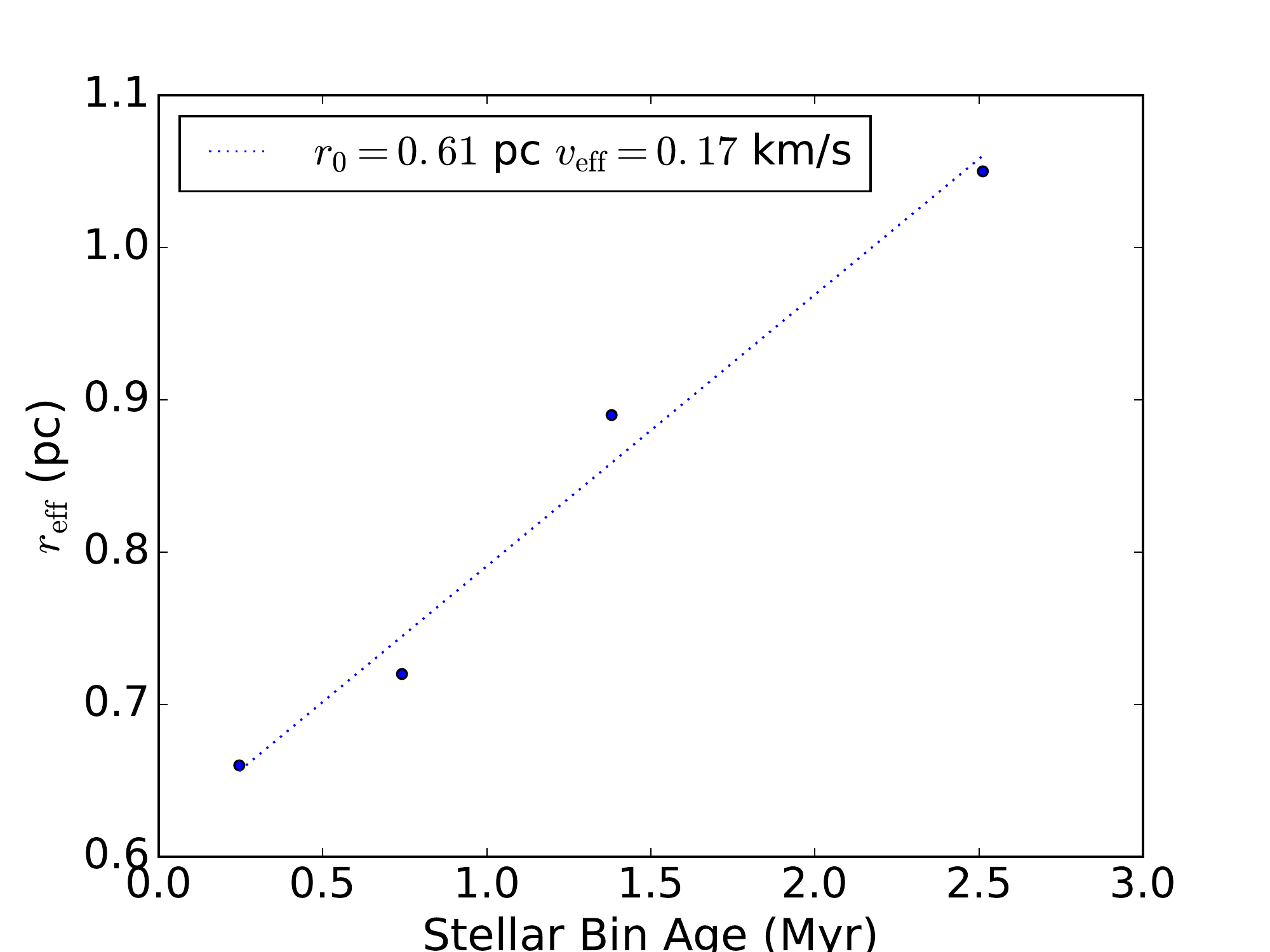}
    \caption{Effective half light radius (blue dots) as a function of age bin for young stars in the ONC. Shown as a thin dotted line is a linear fit given by equation (\ref{eq:reff}) with $r_0 = 0.61$ pc and $v_{\rm eff} = 0.17\,{\rm km\,s}^{-1}$ which is similar to the sound speed and much lower than the turbulent velocity. \label{fig:figure4}}
\end{figure}

In Figure \ref{fig:figure4}, we plot the effective half-light radius $r_{\rm eff}$ as a function of time, $t$, for the mean ages of the the stars in each bin as in Figure \ref{fig:figure 3}.  We fit $r_{\rm eff}(t)$ with a linear expansion of the form,
\begin{eqnarray}
r_{\rm eff} = r_0 + v_{\rm eff} t,\label{eq:reff}
\end{eqnarray}
where $r_0$ is the initial radius of the star cluster and $v_{\rm eff}$ is the effective expansion velocity.  The fits reveal that $r_0 = 0.61$ pc and the expansion velocity is 0.17 km s$^{-1}$.  Equation (\ref{eq:reff}) and Figure \ref{fig:figure4} implies that the stars that form in the ONC over the past 2.5 Myrs formed around a common center and if this region of formation was fixed with an initial size of $r_0$ that the stars that form there are migrating outward with an average speed of $v_{\rm eff}$. This velocity is much lower than the typical turbulent velocity ($v\sim 1\,{\rm km\,s}^{-1}$), which suggests that star formation occurs in a region of converging flows and forms dynamically cold stars.  Coupled with the fact that the region that forms these stars over 2.5 Myrs is the same, it also suggests that structure that formed these stars is quasi-static.

%

\section{Discussion} \label{discussion}

The results of this paper is summarized as follows:
\begin{enumerate}
 \item Star formation is accelerating in MCs.  This is in line with an older observational result \citep{1999ApJ...525..772P,2000ApJ...540..255P,2002ApJ...581.1194P} and with more recent theoretical results \citep{2014MNRAS.439.3420M,2015ApJ...800...49L,2015ApJ...804...44M,2017MNRAS.465.1316M,Murray+18}.  Other groups have also argued for some sort of acceleration to explain the large spread in protostellar ages \citep{2012MNRAS.420.1457H,2012ApJ...751...77Z,2017MNRAS.467.1313V}.
 \item The SFE is superlinear.  For $t_0 \gtrsim 2 t_{\rm ff}$, a superlinear SFE is evident.  There is some evidence that it follows a $t^2$ power law, but this depends on how the initial time $t_0$ is defined.
 \item The star formation in the ONC over the last 5 Myrs is centrally concentrated. When the stars are divided up into different age bins, they are all concentrated to within about 0.5 pc.  Moreover, when distribution of stars broaden with greater age, they can be modeled as a linear expansion with an effective velocity of 0.17 km s$^{-1}$.
\end{enumerate}

The global acceleration of star formation (point i) must be attributed to a global process.  Previously, \citet{2000ApJ...540..255P} argued for global collapse, but such a process would produce stars everywhere as the dense pockets that collapse and form stars would vanish in one local dynamical time.  Hence, while most of the young stars would appear clustered, this would be attributed to having the most recent round of collapse produce most of the stars.  The turbulent collapse model of \citet{2015ApJ...804...44M} would naturally explain the origin of SFE $\propto t^2$ (point ii), but as mentioned this is somewhat sensitive to the definition of $t_0$.  Notably this is also a problem for simulations \citep{2014MNRAS.439.3420M,2015ApJ...800...49L,2015ApJ...806...31G,2015ApJ...804...44M,2017MNRAS.465.1316M,Murray+18}, but here the definition is easier as the formation of the first star in a simulation is a measurable quantity.

We note that several other lines of observational evidence suggest that the star formation rate of molecular clouds accelerate in time.  For instance, it has been noted by \citet{1988ApJ...334L..51M} that there exists a large spread (by a factor of about 300) in the star formation rate among molecular clouds in our galaxy. This has been confirmed by more recent measurements \citep{2009ApJS..181..321E,2010ApJ...724..687L,2011ApJ...729..133M,2016ApJ...833..229L}.  \citet{2012ApJ...745...69K} attributed this disparity to variations in the local free-fall time in clouds, but \citet{2014ApJ...782..114E} found that a large dispersion in the star formation rate still exists even after accounting for the variation in local free-fall times.

\citet{2012ApJ...746...75M} suggested that a time variable, i.e., accelerating, star formation rate could account for the observed scatter.  This suggestion drove much of the numerical \citet{2014MNRAS.439.3420M,2015ApJ...800...49L,2017MNRAS.465.1316M,Murray+18} and analytic \citet{2015ApJ...804...44M} work on star formation in turbulent collapsing molecular clouds.  Observationally, \citet{2016ApJ...833..229L} cross-correlated the molecular cloud catalog produced by \citet{2017ApJ...834...57M} with the star-forming complexes catalog produced by \citet{2012ApJ...752..146L} and showed that the dispersion of the star formation rate varies over 3-4 orders of magnitude (much larger than \citealt{1988ApJ...334L..51M}).  More importantly, they examined many different models to produce these large dispersions and found that a star-forming efficiency of $t^3$ (even larger than the $t^2$ star-forming efficiencies found in simulations) best explains the large variation.

The observation that old and young star formation occurs in the same region in the ONC does bolster the turbulent collapse model.  Here, the model of \citet{2015ApJ...804...44M} predicts that as the density approaches an attractor solution, dense collapsing regions in a turbulent medium can persist for many local dynamical times.  Hence all the star formation would occur in a concentrated well-defined region as opposed to many different regions.  By examining the clustering of stars as a function of age, we have shown that all the stars are centrally concentrated to within about 0.5 pc and that this region is expanding slowly with an effective expansion velocity similar to the cold gas sound speed.  Note that this is different from a global collapse model which does not prescribe how the forming stars are distributed in the the molecular cloud.  In this case, it would be expected that dense star-forming regions do not persist for many local dynamical times and so stars of different ages would cluster differently.

We should note that point (ii) does not strictly endorse the model of \citet{2015ApJ...804...44M}.
For instance, \citet{2006ApJ...641L.121T} argued that the rate of star formation in dense gas should be slow and thus this dense gas can persist for many dynamical times.  How these structures remain in quasi-equilibrium, however, is unclear in this case. Moreover, this picture would presume that the rate of star formation remains constant in contrast with the observed acceleration. However, coupling this quasi-equilibrium gas with some form of global collapse may provide the observed acceleration.

Other models of time-varying star formation are also present in literature and fall either under the aegis of global gravitational collapse \citep{2012MNRAS.420.1457H,2012ApJ...751...77Z,2017MNRAS.467.1313V} or quasi-static evolution \mbox{\citep{1999ApJ...525..772P,2000ApJ...540..255P,2006ApJ...644..355H,2007ApJ...666..281H}}.
For instance, \citet{2006ApJ...644..355H,2007ApJ...666..281H} explored a simplified quasi-equilibrium model of the ONC by modeling it as a collapsing isothermal sphere supported by turbulent pressure and applied the Schmidt law as their model for star formation.  This quasi-equilibrium model does account for much of the observed acceleration and the balance between turbulent pressure and self-gravity and can produce quasi-equilibrium dense structures, but it assumes that molecular clouds begin in equilibrium, which is unclear if they do.

\citet{2012MNRAS.420.1457H} have argued using numerical simulations that the existence of $>5-10$ Myr old stars in nearby molecular clouds is consistent with rapid evolution (i.e., collapse) of a molecular cloud over a few Myrs.  Similarly, \citet{2012ApJ...751...77Z} argue on a semi-analytical basis that the global collapse of molecular clouds sets an initially slow star formation rate that accelerates in time. \citet{2015ApJ...804...44M} is fully consistent with both statements as timescale of the collapse is set by the mean density of the cloud.  The difference is that our theory links the local density and structure velocity of collapsing regions with the SFE and makes a somewhat different prediction for the time dependence compared to \citet{2012ApJ...751...77Z}. Although, in the regime for which comparison to numerical simulations is possible, the predictions are similar.  In addition, the regions of star formation may not be constant, but rather can differ in different epochs, though if the cloud is naturally centrally concentrated, the regions of star formation would also be centrally concentrated \citep{2012MNRAS.420.1457H}.

In the case of the ONC discussed in \S~\ref{ONC}, the star formation is indeed centrally concentrated and persists for many local dynamical times, but the story is not so clear for the other star-forming regions -- Perseus, $\rho$ Ophiuchi, and Taurus.  In our maps (not shown) of Perseus and $\rho$ Ophiuchi, the regions of star formation remain localized across different age bins. For the case of Perseus, the regions of star formation are not centrally concentrated and are spread over $\sim 20$ pc. For $\rho$ Ophiuchi, the regions of star formation are concentrated in a regions of $\sim 2$ pc, but the statistics are much poorer than for the ONC.

\section{Conclusions}\label{sec:conclusions}

We studied the history of star formation in four MCs: Taurus, Perseus, Orion A, and $\rho$ Ophiuchi.  By using published mass and age estimates for each MC, we were able to reconstruct the history of star formation in each cloud in a manner similar to \citet{1999ApJ...525..772P,2000ApJ...540..255P,2002ApJ...581.1194P}.  In agreement with their results, we found that the star formation rate over the last 10 Myrs has been accelerating.  We also find that the star-forming efficiency is broadly consistent with a superlinear star formation rate and some evidence that it follows a
$t^2$ (quadratic in time) power law in line with recent analytic and
numerical studies \citep{2014MNRAS.439.3420M,2015ApJ...800...49L,2015ApJ...806...31G,2015ApJ...804...44M,2017MNRAS.465.1316M,Murray+18}.
In particular, the analytic turbulent collapse model of \citet{2015ApJ...804...44M} naturally produces a $t^2$ power law because the density profile around collapsing regions approach an attractor solution and the infall velocity is proportional to $\sqrt{M_*}$, where $M_*$ is the mass of the central star or star cluster.

Because the density approaches an attractor solution, structures in collapsing regions can persist for far longer than their local dynamical times (\citealt{2015ApJ...804...44M}; see also the simulations of \citealt{2017MNRAS.465.1316M}).  To examine this possibility, we then studied the spatial distribution of star formation in the ONC and found that the the stellar density of stars in different age bins all peak toward a common center, but the distribution broadens with
increasing age. The central region of star formation has persisted for at least 5 Myrs, which is at least an order of magnitude longer than the local free-fall time.  We computed an effective half-light radius, $r_{\rm eff}$, for each age bin and found that this radii can be modeled
as a linear expansion with time with an effective radial expansion
velocity of $v_{\rm eff} \approx 0.17\,{\rm km\,s}^{-1}$, which is
roughly the sound speed of the cold molecular gas.
The spatial distribution of the young stars suggest that
they are formed in centrally concentrated regions that persist for
many local dynamical times.

The global acceleration of the star formation rate has been noted by a number of workers previously observationally \citep{2000ApJ...540..255P,2016ApJ...833..229L}.  The fact that the star formation efficiency scales like $t^2$ and appears to have a common center for all the young stars independent of their ages suggests that some form of turbulent collapse like that proposed by \citet{2015ApJ...804...44M} may be responsible though as we noted competing models \citep{2006ApJ...644..355H,2006ApJ...641L.121T,2007ApJ...666..281H,2012MNRAS.420.1457H,2012ApJ...751...77Z} may also be operating.

\section*{Acknowledgments}

We thank the anonymous reviewer for constructive comments.
SC is supported in part by the Office of Undergraduate Research at the University of Wisconsin-Milwaukee.
PC is supported in part by the NASA ATP
program through NASA grant NNX13AH43G, NSF grant AST-1255469, and the University of Wisconsin-Milwaukee.




\bibliographystyle{mnras}
\bibliography{Bibliography} 

\bsp	
\label{lastpage}
\end{document}